\documentclass[11pt,twoside]{article}

\usepackage{newpasp} 
\usepackage{epsf} 
\usepackage{epsfig}
\usepackage{rotating} 
\usepackage{makeidx} 
\usepackage{minitoc}

\reversemarginpar

\def\rmd{{\rm d}}

\def\edcomment#1{\iffalse\marginpar{\raggedright\sl#1\/}\else\relax\fi}

\begin{document}

\markboth{Boily}{Rotation and Cluster Dynamics}

\title{The Impact of Rotation on Cluster Dynamics}
\author{Christian M. Boily}
\affil{Astronomisches Rechen-Institut, Heidelberg, Germany; \\ and Observatoire Astronomique de Strasbourg, France}

%
% KEEP THE BOX EMPTY FOR THE DATE
% 
% 

% \input{../../../../tex/mymacros.tex} 

\newcommand{\correct}[1]{{ #1}}         % was \sc before 
\newcommand{\tcol}{ t_{\rm col} } 
\newcommand{\tcr}{ t_{\rm cr} } 
\newcommand{\Bv}[1]{ \mbox{\boldmath $ #1$} }

\begin{abstract} 
The evolution of rotating, isolated  clusters of stars up to core-collapse is investigated 
with  n-body numerical codes. The simulations  start off from  
axisymmetric generalisations of King profiles, with added global angular momentum.  
In this contribution we report on results obtained  for two sets of
single-mass cluster simulations. These confirm the more rapid evolution of
even mildly-rotating clusters.  A model is presented with rotational energy
comparable to $\omega$Centauri's; it reaches 
core-collapse in less than half the time required for  non-rotating clusters. 
 \end{abstract}  

\section{Background} 
 Star clusters  
% are  amongst the oldest associations of stars
%  known. Their velocity dispersion $\sigma_{\rm 1d} \sim 1 km/s$ matches up 
% the integrated light (or, mass) as deduced from the scalar virial 
% theorem. Thus they 
are  % pk - erase  
self-gravitating Newtonian systems of choice where to  brew complex 
gravitational dynamics (Meylan \& Heggie 1997 for a review). 
Observations of old, globular, stellar clusters have led to the formulation of spherically symmetric dynamical models
 of equilibria. The most successful and universally studied 
 one-integral  spherical models are the King (1966) profiles.  
However, surveys of up to 100  Milky Way clusters have
 found small but significant departures from spherical symmetry (White 
 \& Shawl 1987): fits to their 
projected isophotes yield ellipticities $<\!\! \epsilon\!\!> \equiv 1 
 - <\!\! a/b\!\!> \approx 0.07 \pm 0.01$.    A study of 173 clusters in M31 
 found  $<\!\!\epsilon\!\!> = 0.09 \pm 0.04$ (Staneva, Spassova \& Golev 1996). 
Observations  of young clusters in the Large 
 Magellanic Cloud revealed  isophotal contours  with ellipticities as large 
as $ \epsilon = 0.3$ (Elson, Fall \& Freeman 1987; Kontizas et al. 1990). This raises  the
 possibility  that   clusters    are   formed     as  strongly     flattened
 structures which then  evolve towards rounder
 configurations (cf. Frenk \& Fall 1982; Boily, Clarke \& Murray 1999; Theis \&
 Spurzem 1999), 
 and brings up important theoretical issues concerning  processes which 
 may  drive this evolution.   

 Rotation stretches any stellar association  along a preferred  axis:   
observations of  the clusters $\omega$Centauri and M13  
have  shown that they are flattened by
rotation (Meylan \& Mayor 1986; Merritt,  Meylan \& 
 Mayor 1997; Lupton, Gunn \& Griffin 1987).  Thus angular 
momentum, measured or possibly % possibly lost .. - pk 
lost during evolution, offers  a way  to account for the
morphology of  clusters. Yet to date there are few evolutionary models of clusters with
 initial angular momentum. 
We have started on  a project to develop three-dimensional dynamical models of
rotating star clusters. In this articles results for 
two n-body models of isolated clusters are presented.  Previous modelling of rotating clusters
 is reviewed  first. 

\section{Gas and Fokker-Planck Models of rotating clusters} 
 Agekian (1958)  considered the effects of 
 angular momentum diffusion on  the equilibria of rotating fluid masses of uniform density.  
In his analysis, concentric spheroids 
rotating about their minor axis become rounder in time when the spheroids have initially an 
ellipticity $\epsilon = 1 - a/b \leq 0.735$, where $a$ and $b$ are the minor and major 
axes. Shapiro \& Marchant (1976)  integrated the equations of motion for this fluid in the 
limit of adiabatic (slow) diffusion of momentum. 
 Angular momentum losses are driven by mass elements  moving in the direction
 of  the stream leaving the system at a rate higher than those moving in the
 opposite direction. The  energy required for escape comes from 
 `heat', attributed to  two-body encounters. Thus,  angular momentum losses are accrued
 over a local two-body relaxation timescale, $\tcol$, which is inversely  proportional to the 
mass density  $(\propto 1/\rho_\star)$. In practice this hinders  applications of
 the results to actual clusters, which show centrally peaked density
 profiles (Meylan \& Heggie 1997). Nevertheless, the framework set by Agekian
 provides a start in  linking rotating bodies and observed (non-rotating) globular clusters. 

With zero rotation, the central region of a cluster evolves towards a cusp in
density during what is known as the gravothermal catastrophe. 
 Does rotation stop the formation of a cusp? 
Hachisu (1979, 1982) discussed the time-evolution  of  self-gravitating cylindrical 
distributions of gas  with angular momentum. He 
predicted a runaway collapse of the central region whenever  angular  momentum is 
expelled faster than a critical rate (see also Lagoute \& Longaretti 1996). Hachisu dubbed this  
the `gravo-gyro catastrophe', by analogy with the non-rotating  case. 
These were until recently the only evolutionary models of rotating
clusters. Two-dimensional orbit-averaged Fokker-Planck methods have
now also been developed to address this issue.  Following Goodman's (1983) approach, 
Einsel \& Spurzem (1999) integrated the Fokker-Planck equation in energy-momentum space  
$[E,J_z]$. Their initial configurations are truncated King models with added bulk motion. 
This velocity field takes the form of a Maxwellian distribution such
that the mean velocity scales  
in proportion to radius away from the centre, then drops off at large
radii. Their adopted axisymmetric distribution function 
(cf. Lupton, Gunn \& Griffin 1987) 

\begin{equation} f(E,J_z) \propto \exp\left( -\beta\Omega_o J_z \right)\cdot  \left[ \exp \left( 
- \beta E \right) - 1 \right] \end{equation} 
where $\beta$ is the inverse square central velocity dispersion and $\Omega_o$ an angular velocity. 
The initial conditions are fixed by specifying  the dimensionless parameters 
\[\omega_o =  \Omega_o/\sqrt{9 G\rho_c/(4\pi)}\ {\rm and}\ W_o\ ,\]
ie, the scales of angular momentum and gravitational 
potential, respectively. The latter is the King parameter. 

In the 2D Fokker-Planck models core-collapse proceeds on a much shorter timescale than in the 
non-rotating case, confirming Hachisu's early intuition. 
% 
% At first angular velocity increases in the central region. 
%  However after 2 to 3 two-body relaxation times ($\tcol$), 
% angular momentum levels off although the   
% velocity  dispersion continues to increase with the central density. 
% This suggests that the `catastrophe'  envisaged  
% by Hachisu did not take place, but rather was a transient phase
% towards asymptotic evolution: 
% 
% $^{\P}$\footnotetext{$^\P$Akiyama \&
% Sugimoto (1989) also  identified transient phases in their cluster models. The
%high level of noise of their evolution tracks leaves open questions
%regarding the quantities involved.}
% Einsel
% \& Spurzem note that the angular velocity of their solutions 
However the central  angular velocity does not 
increase at the high rates expected during the on-set of a 
 gravo-gyro catastrophe; however near the end of core-collapse
 the central  velocity dispersion  
bears the same relation to the central density   
as in  the non-rotating self-similar collapse. 
This leaves open the 
question of what controls the final phase of evolution in these
systems, ie whether or not rotation truly survives up to core-collapse. 
 We chose to  approach this problem using three-dimensional numerical
 integration; the setup is summarised below, followed by results and a
 discussion. 
\section{Basic properties}
\begin{figure}[t] 

\setlength{\unitlength}{1in} 
\begin{picture}(4.,3.5)(0,0) 
        \put(1.5,2.){ \epsfysize=.25in
                  \epsffile[ 10 00 150 20]
                   {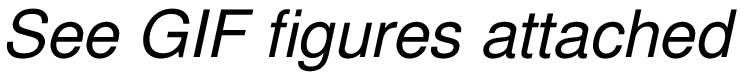} 
                  }
       \put(0.,.5){
\begin{minipage}[t]{\textwidth} \caption{\label{fig:ics} Initial profile of models with
$W_o = 6.0$ and four different values of $\omega_o$. The parameter
       $\eta$ defined in (\ref{eq:eta}) is a measure of anisotropy,
       while $\Omega$ is the angular speed at radius $r$ (in model units). 
       Both quantities are averaged over spherical shells.}  \end{minipage} 
       } 
\end{picture}  
\end{figure}

% 
% The
% by-now classic Ahmad-Cohen (1973) algorithm
% splits the force-field  into
% 'regular' (long timesteps)  and 'irregular' (short timesteps) terms,  according to 
% inter-particle distances. 
%
%
% \section*{Setup of initial conditions}\label{setup}
Self-consistent n-body realisations of the distribution function were
 obtained from the equilibrium Fokker-Planck code {FOPAX}
developed by Christian Einsel.
The models are fully specified  once values are assigned to $(W_o,
\omega_o)$. Figure \ref{fig:ics} illustrates the properties of a set of models of 
10,000 particles with 
$W_o = 6.0$ and four values of $\omega_o$. 
The model clusters rotate about the z-axis and the
equator lies in the x-y plane of a Cartesian coordinate system. 
Rotation causes the cluster in equilibrium  to  flatten down the z-axis and 
this is shown from computing the components of the inertia tensor
$I_{ij}$ 
for a series of twenty concentric spherical shells of equal mass $\rmd M$. 
We define 

\begin{equation} \eta[r_k]  \equiv 1 - \frac{2 \ I_{zz} }{I_{xx} +
I_{yy} }\ ,\ \forall\ {\rm particles\ in\ }\ r_k\, -\, \rmd r < r < r_k\,+\,\rmd r \ . 
\label{eq:eta} \end{equation} 
The parameter $\eta = 0$ when the mass within a shell is
distributed isotropically; $\eta < 0$ (or, $>0$) when the distribution
is anisotropic oblate (or, prolate). For the spherical model $\omega_o = 0$ we
found indeed near-zero values of $\eta$ at all radii. 

Models with rotation have  $\omega_o \neq 0$ and a range of values 
for $\eta$ increasing with it. Note that all models, save one with 
$\omega_o = 0.8$, have  values of $\eta$ compatible with sphericity at the centre. 
 At larger radii, the models are all distinguished from one another. 

Fast-rotating models need be more compact  in order to sustain the
accrued centrifugal force, 

\[ {\rm centrifugal\ force} = \frac{v_\phi^2}{r} = r \Omega^2 \ , \]
which must always  be smaller than the gravitational force, giving the 
condition 

\begin{equation}  \Omega^2 \leq \frac{GM}{r_s^3}\ . \label{eq:centrifugal} \end{equation} 
Thus at constant mass $M$ the system radius $r_s$ must be smaller to
allow for larger angular speed $\Omega$. This is illustrated  on 
 figure \ref{fig:ics}, which displays $\eta$ and $\Omega$ computed from the
 same set of particles. All models show $\Omega$  decreasing
with radius. Note that the curves are consistent with
solid-body rotation in the core-region. Further out $\Omega$ declines
to near-zero, in a trend opposite that of $\eta$. The core remains
roundish despite the large angular speed because the gravity  is
relatively stronger there than near  the edge, and so random motion of the
particles dominate over streaming motion. Overall the fraction of
kinetic energy invested in streaming  motion ranges from 0\%
to 4\%,  14\% and 26\% in increasing order of $\omega_o$. For comparisons, 
 the cluster $\omega$Centauri invests perhaps as much as 22 \% of its kinetic 
energy in rotation (Merritt, Meylan \& Mayor 1997).
% Although it
% was defined as function of the central density $\rho_c$ and angular
% speed $\Omega_o$ only, the parameter $\omega_o$ indeed characterises 
% globally the morphology of the clusters. 

% Put figure 1 here - 

\begin{figure*}[t] 
\setlength{\unitlength}{1in} 
\begin{picture}(4,4)(0,0) 

\put(-.5,-1.68){\epsfig{file=see.figures.ps, height= 0.\textheight,
width=.0\textwidth, angle=-0} 
                  }
        \put(1.5,2.){ \epsfysize=.25in
                  \epsffile[ 10 00 150 20]
                   {see.figures.ps} 
                  }
       \put(.0,0.){ \begin{minipage}{\textwidth} \caption{ \label{fig:corecollapse}
       Time-evolution of the central density (left-most panels), total
       z-angular momentum (middle) 
       and the mean- and core-radii (right-hand panels) for
       runs with $W_o = 6.0$.} \end{minipage} }
\end{picture}  
\end{figure*}

\begin{figure*}[t] 
\setlength{\unitlength}{1in} 
\begin{picture}(3.,4.)(0,0.) 

        \put(1.5,2.){ \epsfysize=.25in
                  \epsffile[ 10 00 150 20]
                   {see.figures.ps} 
                  }

       \put(.0,0.){ \begin{minipage}{\textwidth} \caption{ \label{fig:angmomentum}
       Distribution of specific angular momentum components for three
        time-slots.  See text for explanation 
% f.l.t.r: initial distribution, half-way to, 
%        and at core-collapse. The top set show the z-component; the bottom
%         the x-component.
        of the black squares. } 
       \end{minipage} }
\end{picture}  
\end{figure*}

\section{N-body simulations} 
The code  {NBODY6++} is  an Aarseth-type integration
code based on a Hermite expansion of the variables in time (Aarseth 1999).  
 It has been ported to parallel architecture (Spurzem 2000);  the
 calculations were performed on CRAY computers linked up with MPI library.  
The code treats particles as point-masses and stellar evolution options were
switched off.  
The chain-regularisation algorithm  for hierarchical stellar encounters  as
well as the standard `KS' regularisation 
 (Mikkola \& Aarseth 1998; Aarseth 1999) ensures high-precision integration during close
 interactions. Only simulations with N = 5,000 equal-mass particles will be
 discussed. There are no external tides. 

Figures \ref{fig:corecollapse} \& \ref{fig:angmomentum} illustrate the
time-evolution of the models. The central density, total angular momentum and 
mean and core  radii are plotted as function of time in units of the two-body 
relaxation time $\tcol $ (see Meylan \& Heggie 1997; Casertano
\& Hut 1985). (Note: the lengths were normalised to their initial values.)  
The top panels show evolution for the case of $\omega_o = 0.5$, the bottom set for 
$\omega_o = 0.8$. Looking 
at these diagrams we find an evolution of the central density similar to the
standard case with no rotation: the contraction of the central region leads to 
more close encounters and ejection, hence further contraction ensues, etc,
until $t \simeq 5~\tcol$ when the density peaks sharply, 
indicating core-collapse: at the end of the simulations   
$r_c \approx 0.03$ and $0.015$, respectively, for $\omega_o = 0.5$ and 0.8. At 
constant energy, core-contraction drives the expansion of the outer envelope
and hence the mean radius expands rapidly at core-collapse. Note a subtle
but noticeable difference between the two simulations, namely that the cluster
with initially more rotation evolves faster; this is particularly visible in 
a comparison of radii at fixed time. A more convincing demonstration of the 
fast evolution of such  clusters follows if we recall that in this unit of
time, clusters without rotation reach core-collapse in around $12~\tcol$,
which is more than twice as long. 

To appreciate how many stars might be lost to galactic tides, were a tidal 
field present, we imagine the cluster orbiting the galaxy on a circular
orbit. A tidal radius may then be defined from the initial configuration, by
computing the radius $r_t$  at which the mean density at time $t$ equates 
the initial mean density:  $r_t(t) = <\!\!r\!\!>[0] \times ( M[t]/M[0] )^{1/3}
$. If we label as escapers all stars found outside $2~r_t(t)$, we obtain an
estimate of the number of stars likely to leave the cluster on a timescale 
short compared with $\tcol$; the angular momentum they carry with them is 
deduced from summing up all the momenta of the stars left behind, and comparing 
with the initial value. Implementing this algorithm, we found 
 the run with $\omega = 0.5$ (top panels) would have lost  3.7\% of
the initial 
angular momentum over the time of evolution, but only 1.0\% (51:5000)
of its mass. The second model, with more rotation,  would have lost  0.96\% (48:5000)
of its mass, but only 1.4\% of its total momentum to such escapers. This 
shows how the cluster  redistributes angular momentum efficiently within 
itself, such that the core evolves  towards core-collapse despite 
added rotational support. Evolution in the core is faster, the faster the 
core rotates initially, since the cluster is more compact
(cf. Eq. \ref{eq:centrifugal} and Fig. \ref{fig:ics}) which speeds up
two-body effects. 

Figure \ref{fig:angmomentum} graphs the specific angular momentum of the
$\omega_o = 0.8$ cluster as a function of radius $r$ for three different times. 
For comparison, two components are given: the z-axis component about which
the
cluster rotates; and the x-axis component. Dividing the cluster in ten  
concentric shells, we computed $ \Bv{L} = \Bv{r} \times \Bv{v} $ % bold, vectors - pk 
and summed 
up the momenta in each shell: the result is the series of black squares 
shown on the figure. Initially the (net) z-angular momentum increases steadily 
from the centre, outwards; the symmetry of the figure would make the sum over 
$L_x$'s cancel out, and the black squares have been left out for this
quantity.  Evolution is monotonic, with the net z-momenta inside $r = 2$pc 
decreasing, from which we deduce that an increasing fraction of the momentum 
is transferred to the volume  $> 2$pc. 
Notice on figure \ref{fig:angmomentum} that the stars form a core around $r =
0.5$pc in the final stage of the simulation (right-most panels). 
It is not clear whether this is the result of an 
$m=1$ (lopsided) instability, attributable to the dynamics of the 
system (from a d.f. point of view), or a case of core-wandering, likely 
due to the small number of particles  inside $r = 0.5$pc (Sweatman 1993). 

\section{Conclusion} The faster evolution of clusters with rotation has 
been illustrated with two sample runs. 
The time to core-collapse we found from three-dimensional n-body simulations 
are in  agreement with two-dimensional Fokker-Planck calculations (Einsel \&
Spurzem 1999): 
the collapse time of 5.4 $\tcol$ obtained  
for the $\omega_o = 0.8$ agrees with  the Fokker-Planck solution of 
5.6 $\tcol$ for these parameters. 
 This increases confidence in the results up to core collapse, obtained 
with two different algorithms. 

\acknowledgements Operating 
grants from  the 
Neumann Institute for Computing, J\"ulich, and the Centre for
High-Performance Computing, 
Stuttgart, awarded to R. Spurzem (ARI) are gratefully acknowledged. 
CMB was funded by research grant A/99/49003 awarded by the German DAAD in 1999. 
 Thanks to P. Kroupa for commenting on some aspects of this paper. 

% Start figure captions only here ... 

%
% BEGIN THE REFERENCE LIST WITH \beginrefer
% USE \refer BEFORE THE REFERENCES AND BEGIN A NEW PARAGRAPH AFTER THE 
% REFERENCE !
% DO NOT FORGET TO END THE LIST WITH \endrefer
%

\begin{question}{A.Eckart} What happens to the rotation of the cluster
after core-collapse? \end{question} 

\begin{answer}{C.M.Boily} I stopped my simulations precisely at
core-collapse. What you measure in the envelope depends a lot  on
e.g. the galactic tide, which is absent in these simulations. In the core region
I expect evolution will proceed much as in the standard case without 
rotation, since the cusp is isotropic and carries little net
momentum. More detailed modelling is needed here. \end{answer} 
\end{document}